\documentclass[letterpaper, 10 pt, conference]{ieeeconf}  

\IEEEoverridecommandlockouts                              

\let\labelindent\relax
\usepackage{graphicx} 
\usepackage{xspace}
\usepackage{enumitem,enumerate}
\usepackage{xcolor, color}
\definecolor{lightskyblue}{rgb}{0.53, 0.81, 0.98}
\usepackage{enumitem}
\usepackage{siunitx}

\usepackage[T1]{fontenc}

\usepackage{multirow}
\usepackage{nccmath}
\usepackage{amsfonts}
\usepackage{amsmath}
\usepackage{mathtools}
\usepackage{mathrsfs}
\usepackage{graphicx}
\usepackage{subcaption}
\usepackage{algpseudocode}
\usepackage{booktabs}
\usepackage{xparse}

\usepackage{bm}
\usepackage{tikz}
\usetikzlibrary{positioning, arrows.meta, shapes, calc}
\usetikzlibrary{fit, backgrounds}

\NewDocumentCommand{\basis}{om}{
	\IfNoValueTF{#1}
	{\| \Phi^{#2}\|^2}
	{\langle \Phi^{#1},\phi^{#2} \rangle}
}

\NewDocumentCommand{\pce}{om}{
	\IfNoValueTF{#1}
	{\textsf{#2}}
	{\textsf{#2}^{#1}}
}

\newcommand{\mbb}[1]{\mathbb{#1}}
 
\newcommand{\mcl}[1]{\mathcal{#1}}

\newcommand{\splk}[2]{\mcl{L}^2(\Omega, \mathcal{F}_{#1}, \mu; \mathbb{R}^{#2})}
\newcommand{\lspx}[1]{\mcl{L}^2(\Omega, \mathcal{F}, \mu; \mathbb{R}^{#1})}
\newcommand{\lspr}{\mcl{L}^2(\Omega, \mathcal{F}, \mu; \mathbb{R})}
\newcommand{\lsp}{\mcl{L}^2} 

\newcommand{\diff}{\mathop{}\!\mathrm{d}}

\newcommand{\mean}{\mbb{E}}
\newcommand{\var}{\mbb{V}}

\newcommand{\ini}{_{\text{ini}}}
\newcommand{\tini}{\text{ini}}

\newcommand{\dimx}{{n_x}}
\newcommand{\dimu}{{n_u}}

\newcommand{\dimz}{{n_z}}
\newcommand{\dimy}{{n_y}}

\newcommand{\I}{\mathbb{I}}
\newcommand{\N}{\mathbb{N}}
\newcommand{\R}{\mathbb{R}}

\newcommand{\Xini}{X_{\text{ini}}}

\newcommand{\End}{\hfill $\square$}

\newcommand{\code}[1]{\texttt{#1}}
\newcommand{\packname}{\texttt{PolyOCP.jl}\xspace}

\newtheorem{definition}{Definition}
\newtheorem{remark}{Remark}
\newtheorem{example}{Example}

\newtheorem{assumption}{Assumption}

\title{\LARGE \bf
	\packname\xspace -- A Julia Package for Stochastic OCPs and MPC$^\star$
}

\author{Ruchuan Ou$^{1}$, Learta Januzi$^{1}$, Jonas Schie\ss{}l$^{2}$, Michael H.\ Baumann$^{2}$, Lars Gr\"une$^{2}$ and Timm Faulwasser$^{1}$
\thanks{$^\star$The authors acknowledge funding by the Deutsche Forschungsgemeinschaft (DFG, German Research Foundation) - project number 499435839. LJ was supported by the Deutsche Forschungsgemeinschaft (DFG, German Research Foundation) - project number 527447339.}
\thanks{$^{1}$RO, LJ, and TF are with Institute of Control Systems, Hamburg University of Technology,
		Germany. Email: {\tt\small $\{$ruchuan.ou, learta.januzi$\}$@tuhh.de, timm.faulwasser@ieee.org}}%
\thanks{$^{2}$JS, MHB, and LG are with Mathematical Institute, University of Bayreuth, Germany. Email: {\tt\small $\{$jonas.schiessl, michael.baumann@uni-bayreuth.de, lars.grune$\}$@uni-bayreuth.de}}
}

\begin{document}

\maketitle
\thispagestyle{empty}
\pagestyle{empty}

\begin{abstract}
The consideration of stochastic uncertainty in optimal and predictive control is a well-explored topic. Recently Polynomial Chaos Expansions (PCE) have received considerable attention for problems involving stochastically uncertain system parameters and also for problems with additive stochastic i.i.d. disturbances. While there exist a number of open-source PCE toolboxes, tailored open-source codes for the solution of OCPs involving additive stochastic i.i.d. disturbances in \texttt{julia} are not available. Hence, this paper introduces the toolbox \packname which enables to efficiently solve stochastic OCPs for linear systems  subject to a large class of disturbance distributions. We explain the main mathematical concepts between the PCE transcription of stochastic OCPs and how they are provided in the toolbox. We draw upon two examples to illustrate the functionalities of \packname. 
\end{abstract}

\section{Introduction}
The consideration of stochastic uncertainty in optimal and predictive control is a well-explored topic. Landmark results range from Witsenhausen's counterexample, which shows that in non-Gaussian time-varying linear-quadratic output-feedback settings the optimal policy might be non-linear \cite{Witsenhausen71}, through stochastic extensions of the LQR \cite{Florentin61a,kushner1964maximum} to recent investigations of stochastic MPC \cite{farina13probabilistic,fagiano12nonlinear, hewing18a, mcallister22a, schiessl25stability} and the scenario approach for handling uncertainty~\cite{calafiore06a}.
Indeed, in settings with additive Gaussian uncertainty one may take the classic LQG route \cite{athans71a}. When it comes to non-Gaussian disturbances (of finite expectation and variance) and LTI systems, the propagation of first and second moments is structurally identical to the Gaussian setting. For a detailed comparison of different approaches to uncertainty propagation for dynamic systems, we refer to~\cite{landgraf23a}. However, considering only the first two moments may cause a significant loss of distributional information, e.g., for formulations with non-quadratic cost functions \cite{schiessl25towards} and for chance constraints reformulations involving higher-order moments \cite{pan25dissertation}.
Hence, methods that can work with complete distributional information are of interest. 

To this end, a promising approach is to represent random variables of finite expectation and variance by the coefficients of a series expansion in the basis of the underlying $\mathcal L^2$-space of random variables~\cite{sullivan15introduction}. This approach is often referred to as \textit{Polynomial Chaos Expansion} (PCE); its origins can be traced back to the most cited journal paper of Norbert Wiener \cite{wiener38homogeneous}, and its first applications in systems and control date back to \cite{kim12probabilistic,kim13wiener,kim13generalised} and \cite{fagiano12nonlinear}. While the former papers primarily focus on parametric uncertainty of dynamic systems, the latter one appears to be the first to consider PCE for additive stochastic disturbances. We refer to \cite{mishra24a} for a recent overview of PCE methods in systems and control. 

Indeed there exist a number of toolboxes which facilitate PCE, see Table~\ref{tab:PCE_toolbox_comparison}, and also \cite{mishra24a}, for an overview. \code{PoCET}~\cite{petzke20pocet} addresses Galerkin projection for PCE and parametric uncertainty in dynamic systems. \code{PolyChaos.jl} \cite{muehlpfordt20polychaos} provides a framework for efficient construction of PCE bases using quadrature rules and tensorized computation in \code{julia}.
For instance, \code{PolyChaos.jl} includes an example on how to solve stochastic OCP with parametrically uncertain systems. However, the packages focus on generating the PCE of parameters and do not provide functions specifically for stochastic control.
Interestingly, none of these toolboxes is explicitly conceived to handle additive disturbances acting on a dynamic system. 
Put differently, none of the toolboxes mentioned in Table~\ref{tab:PCE_toolbox_comparison} is tailored to simplify the solution of stochastic Optimal Control Problems (OCP) with additive uncertainty. One reason is that additive i.i.d. disturbance processes acting on dynamic systems require to expand the dimension of the PCE basis as the horizon grows, see, e.g.,~\cite{ou25polynomial} for a PCE-based analysis of LQ optimal control. 

The present paper introduces the toolbox \packname designed to address the gap of existing numerical implementations when it comes to PCE-based solutions of stochastic OCPs \cite{polyocp}. To the end of simplifying the construction of the dynamics of PCE coefficients, we mainly focus on stochastic OCPs with quadratic objectives and additive stochastic uncertainty. The considered OCPs are subject to LTI dynamics with additive non-Gaussian or Gaussian uncertainties and chance constraints. \packname readily allows the inclusion of non-Gaussian models such as Beta or Gamma distributions (or certain polynomial combinations thereof). The user also has substantial freedom to combine i.i.d. and non-i.i.d. disturbance models. To this end, \packname interfaces \texttt{PolyChaos.jl}~\cite{muehlpfordt20polychaos} for specific aspects of the PCE problem transcription and it uses \code{JuMP.jl}~\cite{dunning17jump} as an optimization backend.

The intended use case of \packname is rapid in-silico prototyping of stochastic discrete-time OCPs and corresponding MPC schemes. To the best of the authors' knowledge, \packname is the first \texttt{julia} package that provides a PCE implementation tailored to OCPs with stochastic disturbance processes and to stochastic MPC. The generation of real-time capable code (see e.g. GRAMPC~\cite{Kapernick14a} or CasADi~\cite{Andersson19a} for code generation tools for MPC of deterministic systems) and the consideration of distributional uncertainty via PCE~\cite{pan23distributionally} are beyond the scope of the current \packname implementation. 

The remainder of the paper is structured as follows: Section \ref{sec:problem} provides background on the PCE transcription of stochastic OCPs, while Section~\ref{sec:package} introduces core functionalities of the \packname. Section~\ref{sec:examples} presents selected examples. The paper concludes with  Section \ref{sec:conclusions}.

\begin{table*}[t]
	\caption{Overview of selected PCE toolboxes}
	\label{tab:PCE_toolbox_comparison}
	\centering
	\renewcommand{\arraystretch}{1.15}
	\begin{tabular*}{0.95\textwidth}{@{\extracolsep{\fill}} llll}
		\hline
		Toolbox & Language & Features & License \\ 
		\hline
		\packname & Julia & Uncertainty propagation for stochastic LTI systems and OCPs& MIT \\[2pt]
		
		PoCET~\cite{petzke20pocet} & Matlab & Propagation of stochastic parameters for dynamic systems; focus on Galerkin projection & EUPL-1.2 
        \\[2pt]
		
		PolyChaos.jl~\cite{muehlpfordt20polychaos} & Julia & Construction of orthogonal polynomial bases \& quadratures for arbitrary $\mathcal L^2$ distribution & MIT \\[2pt]

        Chaospy~\cite{feinberg15chaospy} & Python & PCE construction via non-intrusive Galerkin projection, uncertainty propagation & MIT \\[2pt]
        
		UQLab~\cite{marelli14uqlab} & Matlab & Adaptive sparse PCEs surrogate models for moment and sensitivity analysis& BSD 3-clause \\[2pt]
		
		OpenTURNS~\cite{baudin17openturns} & Python/C++ & Metamodeling of uncertainty propagation, efficient sampling \& analytical approaches& LGPL \\[2pt]
		
		Dakota~\cite{adams23dakota} & C++ & Approximation of uncertainty propagation, moment and sensitivity analysis & LGPL \\
		\hline\\
	\end{tabular*}
\end{table*}

\section{Problem Formulation}\label{sec:problem}
Next, we provide some background on PCE and stochastic OCPs and introduce the problem formulation.

\subsection{Polynomial chaos expansions}
The key idea of PCE is that any $\lsp$ random variable can be expressed in a suitable orthogonal polynomial basis. Consider a univariate orthogonal polynomial basis $\{\phi^j(\xi)\}_{j=0}^{\infty}$ that spans the space $\mcl{L}^2(\Omega, \mathcal{F}, \mu; \mathbb{R})$, where $\phi^j(\xi)$ is a polynomial of degree $j$\footnote{In general, one may also consider different indexing schemes where $j$ does not correspond to the degree of $\phi^j$.}, $\xi\in \lsp(\Omega, \mathcal{F}, \mu; \Xi)$ is the argument of polynomials. Here, $\Omega$ is the sample space, $\mathcal{F}$ a $\sigma$-algebra, $\mu$ the considered probability measure. Note that $\xi:\Omega\to\Xi$ is viewed as a function of the outcome $\omega$ and hence $\phi^j(\xi(\omega)) = \phi^j\circ\xi(\omega)$. The orthogonality of $\{\phi^j(\xi)\}_{j=0}^{\infty}$ ensures that
\begin{equation} \label{eq:orthogonality}
	\hspace{-1mm}\langle \phi^i(\xi),\phi^j(\xi) \rangle {=} \int_{\Omega} \hspace{-1mm}\phi^i(\xi) \phi^j(\xi) \diff \mu {=} \delta^{ij}\langle \phi^j(\xi),\phi^j(\xi) \rangle
\end{equation}
holds, where $\delta^{ij}$ denotes the Kronecker delta. The first polynomial $\phi^0(\xi)$ is of degree 0 and thus always chosen to be $\phi^0(\xi) = 1$. Hence, the orthogonality~\eqref{eq:orthogonality} gives that for all other basis dimensions $j>0$, we have $\mean[\phi^j(\xi)]=\int_{\Omega} \phi^j(\xi) \diff \mu = \langle \phi^j(\xi),\phi^0(\xi)\rangle=0$.

The PCE of a real-valued random variable $Z \in \lsp(\Omega, \mathcal{F}, \mu; \R)$ with respect to the basis $\{\phi^j(\xi)\}_{j=0}^{\infty}$ is 
\[
Z(\omega) = \sum_{j=0}^{\infty}\pce{z}^j \phi^j(\xi(\omega)) \quad \text{with} \quad \pce{z}^j = \frac{\big\langle Z(\omega), \phi^j(\xi(\omega)) \big\rangle}{\langle \phi^j(\xi),\phi^j(\xi) \rangle},
\]
where $\pce{z}^j\in \R$ is referred to as the $j$-th PCE coefficient. For the sake of readability, we omit the arguments $\xi(\omega)$, $\omega$ and use the shorthand $\lsp(\R) \coloneqq \lspr$ whenever there is no ambiguity. The first two moments of $Z$ thus can be efficiently computed from its PCE coefficients as
\begin{equation}\label{eq:moments}
	\mean[Z] = \pce{z}^0, \quad \var[Z] = \sum\nolimits_{j=1}^{\infty}(\pce{z}^j)^2\langle \phi^j,\phi^j \rangle.
\end{equation}

In numerical implementations the infinite-dimensional expansions have to be truncated after a finite number of terms. This may lead to truncation errors
\[
\Delta Z(L) = Z -  \sum\nolimits_{j=0}^{L-1}\pce{z}^j \phi^j,
\]
where $L\in\N^\infty\coloneqq \N^+\cup\{\infty\}$ is the PCE dimension. For $L\to\infty$, the truncation error satisfies $\displaystyle \lim_{L\to\infty}\|\Delta Z(L)\|=0$ \cite{cameron47orthogonal, ernst12convergence}. 

\begin{definition}[Exact PCE representation] \label{def:exactPCE}
	The PCE of a random variable $Z \in\lsp(\R)$ is said to be exact with finite dimension $L\in\N$ if $ Z -  \sum_{j=0}^{L-1}\pce{z}^j \phi^j=0$.\End
\end{definition}
The correspondence between Wiener-Askey polynomial families and their underlying continuous stochastic arguments is summarized in Table~\ref{tab:PCEBasis}, which serves as a guideline for selecting the appropriate polynomial basis for random variables with canonical distributions \cite{xiu02wiener}. The Dirac measure, representing a deterministic variable with basis $\phi = 1$, is also included to foster modelling freedom. Furthermore, Table~\ref{tab:PCEBasis} lists the first two nontrivial polynomials (beyond $\phi^0 = 1$) for each distribution type, along with their normalized counterparts $\psi^j$ satisfying $\langle \psi^j, \psi^j \rangle = 1$, obtained via
\[
\psi(\xi) = \frac{\phi^j(\xi)}{\sqrt{\langle \phi^j(\xi),\phi^j(\xi) \rangle}}.
\]
\begin{remark}[Generic affine PCE series] 
	Given an $\mcl{L}^2$ random variable with known distribution, the key to constructing an exact finite-dimensional PCE is the appropriate choice of basis functions. For some widely used distributions, the appropriate choice of polynomial bases and the corresponding affine PCE expansions are summarized in Table~\ref{tab:PCEBasis} \cite{xiu02wiener}. Additionally, a generic (non-orthonormal but orthogonal) basis choice for any random variable $Z \in \lsp(\R)$ is $\phi^0 = 1$ and $\phi^1 = Z-\mean[Z]$, which implies the exact and finite PCE $\pce{z}^0 = \mean[Z]$ and $\pce{z}^1 = 1$. \End
\end{remark}

\begin{table*}[t]
	\caption{Polynomial bases for random variables $Z$ following canonical distributions}
	\label{tab:PCEBasis}
	\centering
	\renewcommand{\arraystretch}{1.35}
    \scalebox{1.0}{
	\begin{tabular*}{0.95\textwidth}{@{\extracolsep{\fill}} llllll}
		\hline
		Distribution of $Z$ & Distribution of $\xi$ & Basis & Orthogonal Basis \(\phi^j\) & Orthonormal Basis \(\psi^j\) & Affine PCE\\
		\hline
        Dirac: \(\delta_x(c)\)  & \( \delta_x(0)\) & Constant 
        & $\phi^0=1$ 
        & $\psi^0=\phi^0$ 
        & \(Z = c\cdot\phi^0 = c\cdot \psi^0\) \\
        
		Gaussian: \(\mathcal{N}(\mu,\sigma^2)\) & \(\mcl{N}(0,1)\) & Hermite  
        & \(\phi^1(\xi)=\xi\) 
        & \(\psi^1=\phi^1\)  
        & \(Z =  \mu\cdot\phi^0 + \sigma\cdot\phi^1\)\\
        
		& & 
        & \(\phi^2(\xi)=\xi^2-1\) 
        & \(\psi^2(\xi)=\phi^2/\sqrt{2}\)  
        & \(\phantom{Z}=\mu\cdot\psi^0 + \sigma\cdot\psi^1\)\\[3pt]
		
		Uniform:  \(\mathcal{U}(a,b)\)  & \(\mcl{U}(0,1)\) & Legendre 
        & \(\phi^1(\xi)=\xi-\tfrac{1}{2}\) 
        & \(\psi^1=2\sqrt{3}\phi^1\)  
        & \(Z = \frac{a+b}{2}\cdot\phi^0 + (b{-}a)\cdot\phi^1\) \\
        
		& & 
        & \(\phi^2(\xi)=\xi^2-\xi+\tfrac16\) 
        & \(\psi^2=6\sqrt{5}\phi^2\)  
        & \(\phantom{Z} = \frac{a+b}{2}\cdot\psi^0 + \frac{b{-}a}{2\sqrt{3}}\cdot\psi^1\)\\[3pt]
		
		Beta: \(\mcl{B}(\alpha{=}2,\beta{=}2)\) & \(\mcl{B}(\alpha,\beta)\) 
        & Jacobi
        & \(\phi^1(\xi)=\xi-\tfrac{1}{2}\) 
        & \(\psi^1=2\sqrt{5}\phi^1\)  
        & \(Z = \frac12 \cdot\phi^0 + 1\cdot\phi^1\) \\
        
		& & 
        & \(\phi^2(\xi)=\xi^2-\xi+\tfrac{1}{5}\) 
        & \(\psi^2=5\sqrt{14}\phi^2\)  
        & \(\phantom{Z} = \frac12 \cdot\psi^0 + \frac{1}{2\sqrt{5}}\cdot\psi^1\)\\[3pt]
		
		Gamma: \(\Gamma(\alpha{=}1,\beta{=}1)\)  & \(\Gamma(\alpha,\beta)\) 
        & Laguerre
        & \(\phi^1(\xi)=\xi-1\) 
        & \(\psi^1=\phi^1\)  
        & \(Z = 1\cdot \phi^0 + 1\cdot\phi^1\) \\
        
		& & 
        & \(\phi^2(\xi)=\xi^2-4\xi+2\) 
        & \(\psi^2=\phi^2/2\) 
        & \(\phantom{Z}= 1\cdot \psi^0 + 1\cdot\psi^1\)\\
		\hline
	\end{tabular*}
    }
\end{table*}

To obtain the PCE of a vector-valued random variable, i.e. random vector $Z = \begin{bmatrix} Z_1 & Z_2 & \cdots & Z_{\dimz} \end{bmatrix}^\top\in\lsp(\R^{\dimz})$, where $Z_i\in \lsp(\R)$ denotes the $i$-the element, one needs to construct a multivariate basis from the basis of its elements. Let the PCE basis of each element $Z_i$ be $\{\phi_i^j(\xi_i)\}_{j=0}^\infty$, then the corresponding multivariate basis $\{\phi(\xi)\}_{j=0}^\infty$ with $\xi = \begin{bmatrix} \xi_1 & \xi_2 & \cdots & \xi_{\dimz} \end{bmatrix}^\top$ reads
\begin{equation}
\begin{split}
      	\{\phi(\xi)\}_{j=0}^\infty  &= \{1\}\cup \Phi^1(\xi) \cup \Phi^2(\xi) \cup \cdots \cup \Phi^\infty(\xi),  \label{eq:multibasis}\\
     \text{with}\quad \Phi^p(\xi) &\coloneqq \bigg\{ \prod_{i=0}^{\dimz}\phi_{i}^{j_i}(\xi_{i}) \,\Big|\,j_i\in\I_{[0,p]},\,\sum_{i=1}^{\dimz}j_i = p\bigg\},
\end{split}
\end{equation}
which is the Cartesian product of individual univariate bases $\{\phi_i^j(\xi_i)\}_{j=0}^\infty$, $i\in\I_{[1,\dimz]}$.
$\Phi^p(\xi)$ is a unified representation of multivariate orthogonal polynomials of degree $p$, where the first two sets are
\begin{align*}
	\Phi^1(\xi) &\coloneqq \{\phi_1^1(\xi_1),\phi_1^1(\xi_2),\ldots, \phi_{\dimz}^1(\xi_{\dimz})\},\\
	\Phi^2(\xi)&\coloneqq \Big\{\phi_1^2(\xi_1),\ldots,\phi_{\dimz}^2(\xi_{\dimz}),\\
 &\hspace{6mm}\phi_1^1(\xi_1)\cdot\phi_2^1(\xi_2),\ldots,\phi_{\dimz-1}^1(\xi_{\dimz-1})\cdot\phi_{\dimz}^1(\xi_{\dimz})\Big\}.
\end{align*}
The number of total terms of such an $\dimz$-variate polynomial basis up to degree $p$ is $(\dimz+p)!/(\dimz!p!)$. Importantly, when $Z_1$,\ldots,$Z_{\dimz}$ are independent random variables, the multivariate basis can be simplified to the union of individual bases of all components, i.e. $\{\phi(\xi)\}_{j=0}^\infty = \cup_{i=1}^{\dimz}\{\phi_i^j(\xi_i)\}_{j=0}^\infty$.
By applying PCE component-wise to the multivariate basis, one obtains the $j$-th PCE coefficient of $Z$ as $\pce{z}^{j} =\begin{bmatrix} \pce{z}_1^j & \pce{z}_2^j & \cdots & \pce{z}_{\dimz}^j \end{bmatrix}^\top$, where $\pce{z}_i^j$ is the $j$-th PCE coefficient of $Z^i$, $\forall i\in\I_{[1,n_z]}$. We illustrate the procedure of generating finite-dimensional multivariate basis via the following two examples, each with a maximum polynomial degree of one.
\begin{example}[Independent components in random vector]\label{example: PCE1}
Consider $Z = \begin{bmatrix} Z_1 & Z_2 \end{bmatrix}^\top\in \lsp(\R^2)$ with $Z_1\sim\mcl{U}(a,b)$ and $Z_2\sim\mcl{N}(\mu,\sigma^2)$, where $Z_1$ and $Z_2$ are independent.
As Table~\ref{tab:PCEBasis} suggests, the exact PCE representations of $Z_1$ and $Z_2$ read
\begin{alignat*}{3}
 	Z_1 & = \frac{a+b}{2}\cdot 1 &&+ (b-a)\cdot\phi_1^1(\xi_1)\quad&&\text{with}\quad \phi_1^1(\xi_1) = \xi_1-\frac12,\\
 	Z_2 & = \mu\cdot 1 &&+ \sigma \cdot\phi_2^1(\xi_2)\quad&&\text{with}\quad \phi_2^1(\xi_2) = \xi_2,
\end{alignat*}
where the stochastic arguments $\xi_1\sim\mcl{U}(0,1)$, $\xi_2\sim\mcl{N}(0,1)$ are independent. By constructing a multivariate basis $\{1, \phi_1^1(\xi_1), \phi_2^1(\xi_2)\}$, the PCE of Z reads
\begin{equation*}
 	Z = \begin{bmatrix} \frac{a+b}{2} \\ \mu \end{bmatrix} \cdot 1 + 
 	 \begin{bmatrix} b-a \\ 0 \end{bmatrix} \cdot \phi_1^1(\xi_1) + 
	  \begin{bmatrix} 0 \\ \sigma \end{bmatrix} \cdot \phi_2^1(\xi_2). \tag*{$\square$} \vspace{2mm}
\end{equation*}
\end{example}
\begin{example}[Multivariate Gaussian] \label{example:PCE2}
	Consider a multivariate Gaussian distribution of a $\dimz$-dimensional random vector $Z = \begin{bmatrix} Z_1 & Z_2 & \cdots & Z_{\dimz} \end{bmatrix}^\top\in\lsp(\R^{\dimz})$ following the distribution $Z\sim\mcl{N}(\mu,\Sigma)$. Then the exact PCE representation directly follows as
	\[
		Z = \mu\cdot 1 + \mathrm{Chol}(\Sigma)\cdot\begin{bmatrix}
			\phi_1^1(\xi_1) & \phi_2^1(\xi_2) & \cdots & \phi_{\dimz}^1(\xi_{\dimz})
		\end{bmatrix}^\top,
	\]
	where $\mathrm{Chol}(\Sigma)\in\R^{\dimz\times\dimz}$ denotes a lower-triangular matrix obtained by Cholesky decomposition of $\Sigma$ \cite{pan23distributionally}. The basis functions are chosen as $\phi_i^1(\xi_i)=\xi_i$ with independent stochastic arguments $\xi_i\sim\mcl{N}(0,1)$ for all $i\in\I_{[1,\dimz]}$.\End
\end{example}
Note that the multivariate bases in the above examples preserve orthogonality because the stochastic arguments are independent random variables. The multivariate Gaussian distribution in Example~\ref{example:PCE2} allows addressing distributionally robust uncertainty propagation within the PCE framework~\cite{pan23distributionally}.

\subsection{Uncertainty propagation}
PCE originated in \cite{wiener38homogeneous} for the parameterization and approximation of random variables and was later introduced to solve stochastic differential equations by \cite{xiu02wiener}. Under suitable assumptions, PCE enables exact uncertainty propagation. Consider an explicit map
\[
	f:~\lsp(\R^{\dimz})\to\lsp(\R^{\dimy}),\quad Z \mapsto Y=f(Z)
\]
and let the PCEs of $Z$ and $Y$ be $Z = \sum_{j=0}^{\infty}\pce{z}^j \phi^j$ and $Y = \sum_{j=0}^{\infty}\pce{y}^j \phi^j$, respectively. Assuming the PCE coefficients of $Z$ are known, the PCE coefficients of $Y$ can be computed via Galerkin projection as follows
\begin{itemize}
\item[(i)] Substitute the PCE expansions of $Z$ and $Y$ in the map $f$, which yields $\sum_{i=0}^{\infty}\pce{y}^i \phi^i = f(\sum_{i=0}^{\infty}\pce{z}^i \phi^i)$
\item[(ii)] Project the map onto the polynomial basis $\phi^j$, $j\in\N^\infty$:
\[
	\left\langle \sum\nolimits_{i=0}^{\infty}\pce{y}^i\phi^i,\phi^j \right\rangle = \left\langle f\big(\sum\nolimits_{i=0}^{\infty}\pce{z}^i \phi^i \big),\phi^j\right \rangle
\]
\item[(iii)] Use the orthogonality property~\eqref{eq:orthogonality} and obtain the coefficients $\pce{y}^j$ via 
\begin{equation}\label{eq:galerkin}
	\pce{y}^j = \frac{ \left\langle f\big(\sum\nolimits_{i=0}^{\infty}\pce{z}^i \phi^i \big),\phi^j\right\rangle}{ \langle \phi^j,\phi^j\rangle}.
\end{equation}
\end{itemize}
\begin{example}[Nonlinear map] \label{example:nonlinear}
	Consider the map $Y= f(Z) = (Z_1{+}Z_2)^2$, where $Z$ is the random vector from Example~\ref{example: PCE1}. Substituting $Z$ with its PCE yields
	\begin{equation*}
		Y = \left(\frac{a+b}{2}+u+(b-a)\cdot\phi_1^1(\xi_1) + \sigma\cdot\phi_2^1(\xi_2)\right)^2.
	\end{equation*}
	It can be observed that the PCE of $Y$ includes polynomials up to degree two, and the coefficient associated with the cross term $\phi_1^1(\xi_1)\cdot\phi_2^1(\xi_2)$ is non-zero. Therefore, a 2-variate polynomial basis up to degree 2 as \eqref{eq:multibasis} suggests is required to exactly represent $Y$.\End
\end{example}
When a linear or affine map $f$ is considered, however, the uncertainty propagation in the PCE framework is significantly easier. Let $Y=f(Z) = AZ$ with $A\in\R^{\dimy\times\dimz}$. Using Galerkin projection \eqref{eq:galerkin} and the orthogonality relation \eqref{eq:orthogonality}, it follows that
\begin{equation} \label{eq:galerkin_linear}
	\pce{y}^j = \frac{ \left\langle\sum\nolimits_{i=0}^{\infty}A\pce{z}^i \phi^i,\phi^j \right \rangle}{ \langle \phi^j,\phi^j\rangle} = A\pce{z}^j.
\end{equation}
Importantly, \eqref{eq:galerkin_linear} indicates that $Y$ can be exactly represented in the basis of $Z$ and thus no new terms are introduced. This is in contrast to nonlinear maps, cf. Example~\ref{example:nonlinear}.

\subsection{Stochastic optimal control}
Next we present the procedure for reformulation of a stochastic OCP such that it can be efficiently solved within the PCE framework. Consider a stochastic discrete-time LTI system
\begin{equation} \label{eq:sys}
	X(k+1) = AX(k) + BU(k) + EW(k), \quad X(0) = X\ini
\end{equation}
with state $X(k) \in \splk{k}{n_x}$ and process disturbance  $W(k) \in \lspx{n_w}$. The probability distributions of the disturbance $W(k)$, $k\in \N$ and the initial condition $X\ini\in \splk{0}{n_x}$ are assumed to be known and $W(k)$, $k\in \N$ are \textit{i.i.d.} random variables. 

In the filtered probability space $(\Omega, \mcl F, (\mcl F_k)_{k\in \N}, \mu)$, the $\sigma$-algebra contains all available historical information, i.e.,
$\mcl F_0 \subseteq \mcl F_1 \subseteq ...  \subseteq \mcl F$.
Let $(\mcl F_k)_{k\in \N}$ be the smallest filtration that the stochastic process $X$ is adapted to, i.e., $ \mcl F_k = \sigma(X(i),i\leq k)$, where $\sigma(X_i,i\leq k)$ denotes the $\sigma$-algebra generated by $X(i),i\leq k$. Then, $U(k)$ is modeled as a stochastic process which is adapted to the filtration $\mathcal{F}_k$, i.e. $U(k) \in \splk{k}{\dimu}$. This imposes a causality constraint on $U(k)$, i.e., $U(k)$ depends only on $X(i)$, $i\leq k$ up to time step $k$. Thus, $U(k)$ may only depend on past disturbances $W(i)$, $i< k$. For more details on filtrations we refer to \cite{fristedt13modern}.

Given the initial condition $X(0)=X\ini$ and random variables $W(k),~k\in \I_{[0,N-1]}$, we consider the following stochastic OCP with horizon $N \in \N^+$,
\begin{subequations} \label{eq:stochasticOCP}
\begin{align}
\min_{X,U} \quad  &\mean \left[ \sum_{k=0}^{N-1} \|X(k)\|_Q^2 +\|U(k)\|_R^2 + \|X(N)\|_{Q_N}^2\right] \label{eq:OCP_obj}\\
\text{s. t. } \quad  &\eqref{eq:sys},\,k \in \I_{[0,N-1]}, \\
&\mbb P [z_{\min}\leq Z(k) \leq  z_{\max} ] \geq 1 - \varepsilon_z,\,k\in\I_{[0,N^\prime]}\label{eq:OCPchance_Z},
\end{align}
\end{subequations}
where $Q$, $Q_N\succeq0$, $R\succ0$, $\|X(k)\|_Q^2\coloneqq X^\top(k) Q X(k)\in\lsp(\R)$, and $(z,Z,N^\prime)\in\{(x,X,N),\,(u,U,N-1)\}$. Note that we have $k\in\I_{[0,N]}$ for the state constraints, while $k\in\I_{[0,N-1]}$ for the input constraints, reflecting that the predicted state trajectory is one step longer than the control trajectory.

Here we consider the chance constraints \eqref{eq:OCPchance_Z} individually imposed on the components of $X$ and $U$, where $\varepsilon$ denotes the probability that the constraint is violated.
To obtain the PCE reformulated OCP, we make the following assumption.
\begin{assumption}[Exact PCEs for $X\ini$ and  $W_k$]\label{ass:exactIniW}
	The initial condition $X\ini$ and all i.i.d. disturbances $W_{k}$, $k\in \I_{[0,N-1]}$ admit exact PCEs with $L\ini$ terms and $L_w$ terms of orthonormal bases, respectively. That is, $X\ini = \textstyle{\sum_{i=0}^{L\ini-1}} \pce{x}\ini^i \psi\ini^i(\xi\ini)$ and $W_k = \textstyle{\sum_{n=0}^{L_w-1}} \pce{w}_k^n \psi_w^n(\xi_k)$ for $k\in \I_{[0,N-1]}$,
    where $\xi_k$ are i.i.d. stochastic arguments.
\end{assumption}
In the above assumption, each $\xi_k$, $k\in\I_{[0,N-1]}$ corresponds to the disturbance $W(k)$ at time step $k$. In the PCEs of i.i.d. disturbances, the identical distribution is expressed via the shared algebraic structure of the basis functions°$\psi_w$, while the stochastic independence is modeled by the use of different arguments $\xi_k$. Similar to the multivariate basis illustrated in Example~\ref{example: PCE1}, we construct a joint orthonormal basis $\Psi=\{\psi^j(\xi)\}_{j=0}^{L-1}$ with $\xi=\begin{bmatrix}\xi\ini & \xi_1 & \cdots \xi_{N-1}\end{bmatrix}^\top$ for $X\ini$ and $W(k)$, $k\in\I_{[0,N-1]}$ as
\begin{align} \label{eq:jointbasis}
&\Psi \coloneqq \{\psi\ini^i(\xi\ini)\}_{i=0}^{L\ini-1}\bigcup\left(\bigcup_{k=0}^{N-1} \{ \psi_w^n(\xi_k)\}_{n=0}^{L_w-1} \right)\\
= &\Big\{ 1,  \psi\ini^1(\xi\ini),...,\psi\ini^{L\ini-1}(\xi\ini), \psi_w^1(\xi_0),...,\psi_w^{L_w-1}(\xi_0),\nonumber\\ 
&\hspace{3cm}...,\psi_w^1(\xi_{N-1}),...,\psi_w^{L_w-1}(\xi_{N-1})\Big\}. \nonumber
\end{align}
with a total of $L = L\ini+N(L_w-1)$ terms, i.e., it grows linearly with the horizon $N$.
Sufficient conditions ensuring that the optimal solution of~\eqref{eq:stochasticOCP}, i.e. $\{X^\star(k)\}_{k=0}^N$ and $\{U^\star(k)\}_{k=0}^{N-1}$, admits exact PCEs in the joint basis $\Psi$ from~\eqref{eq:jointbasis} have been analyzed in~\cite{pan23stochastic}.

By replacing all the random variables in~\eqref{eq:sys} with their PCEs using the joint basis $\Psi$, e.g. $X(k) = \sum_{j=0}^{L-1}\pce{x}^j(k)\psi^j$, the resulting system for PCE coefficients satisfies
\begin{equation} \label{eq:sysPCE}
    \pce{x}^j(k+1) = A\pce{x}^j(k) +B\pce{u}^j(k)+ E\pce{w}^j(k),\quad \pce{x}^j(0) = \pce{x}\ini^j
\end{equation}
for all $j\in\I_{[0,L-1]}$.
Moreover, the orthonormality of basis functions, i.e. $\langle\psi^i,\psi^j\rangle=\delta^{ij}$ for $i,j\in\I_{[0,L-1]}$, implies that
\begin{align*}
    &\mean\left[ \|X(k)\|_Q^2\right] =  \mean\Bigg[\bigg(\sum_{j=0}^{L-1}\pce{x}^{j,\top}(k)\psi^j\bigg) Q \bigg( \sum_{j=0}^{L-1}\pce{x}^j(k)\psi^j\Big)\Bigg]\\
    = &\sum_{i=0}^{L-1}\sum_{j=0}^{L-1} \pce{x}^{i,\top}(k)Q\pce{x}^j(k)\delta^{ij} = \sum_{j=0}^{L-1} \|\pce{x}^{j}(k)\|_Q^2.
\end{align*}

Let Assumption~\ref{ass:exactIniW} hold and consider the joint basis constructed as~\eqref{eq:jointbasis}. Then the PCE reformulation of~\ref{eq:stochasticOCP} reads:
\begin{subequations}\label{eq:PCE_SOCP}
	\begin{equation}
			\min_{\substack{
				j \in \I_{[0,L-1]} \\  \pce{x}^{j}\in \R^{\dimx}\\ \pce{u}^{j} \in \R^{\dimu}
		}}
		\sum_{j=0}^{L-1}  \bigg(\sum_{k=0}^{N-1}\Big(\| \pce{x}^j(k)\|^2_Q  {+}\|\pce{u}^j(k)\|^2_R\Big)  {+}\| \pce{x}^j(N)\|^2_{Q_N}\bigg) \label{eq:OCP_obj_PCE} \vspace*{-0.1cm}
	\end{equation}
\begin{align}
	\text{s. t. } \quad \forall j& \in \I_{[0,L-1]}, \, k \in \I_{[0,N-1]},  \nonumber\\
	\pce{x}^j(k+1)&{=}  A \pce{x}^j(k){+}B\pce{u}^j(k) 
	{+}E\pce{w}^j(k),\,   \pce{x}^j(0){=}\pce{x}\ini^{j}, \label{eq:OCP_dyn_PCE}\\
	\pce{z}^0(k^\prime)   &\pm \gamma(\varepsilon_z)\sqrt{\sum_{j=1}^{L-1} {(\pce{z}^j(k^\prime))}^2} \in  [z_{\min},z_{\max}], \label{eq:OCP_chance_PCE}\\
	\pce{u}^{j'}(k)&= 0,~ \forall j'\in \I_{[L_{\text{ini}}+k(L_w-1)+1,L-1]},\label{eq:OCP_PCE_causality}
\end{align} 
\end{subequations}
where $k^\prime\in\I_{[0,N]}$ for $\pce{z}=\pce{x}$ and $k^\prime\in\I_{[0,N-1]}$ for $\pce{z}=\pce{u}$ in~\eqref{eq:OCP_chance_PCE}. The causality constraint~\eqref{eq:OCP_PCE_causality} follows from the filtration condition, i.e. from $U(k) \in \splk{k}{\dimu}$~\cite{pan23stochastic}.

Using Cantelli's inequality, a conservative component-wise approximation of the chance constraint~\eqref{eq:OCPchance_Z} is 
\[
    \mean[Z] \pm \gamma(\varepsilon_z) \sqrt{\var[Z]}\in[z_{\min},z_{\max}] \; \text{with} \; \gamma(\varepsilon_z) = \sqrt{\frac{2-\varepsilon_z}{\varepsilon_z}},
\]
see \cite{calafiore06a, farina13probabilistic} for details of this reformulation. By calculating the first two moments of $Z$ from its PCE coefficients as indicated in \eqref{eq:moments}, the convex second-order cone constraint~\eqref{eq:OCP_chance_PCE} follows. In particular, when $Z$ is Gaussian distributed, the above approximation is exact if $\gamma$ is selected according to the standard Gaussian quantile function. Moreover, if $z_{\min}=-\infty$ or $z_{\max}=\infty$, \eqref{eq:OCPchance_Z} reduces to a one-sided chance constraint, in which case $\gamma(\varepsilon_z)=\sqrt{(1-\varepsilon_z)/\varepsilon_z}$.

By solving the reformulated stochastic OCP~\eqref{eq:PCE_SOCP}, one obtains the PCE expansion of the optimal input $U^\star(k;\omega)=\sum_{j=0}^{L-1}\pce{u}^j(k)\psi^j(\xi(\omega))$. Note that the realizations of the basis functions $\psi(\xi(\omega))$, $j\in\I_{[0,L-1]}$ are determined by the disturbance realizations $W(k;\omega)$, $k\in\I_{[0,N-1]}$. Hence, the optimal input $U^\star$ obtained by solving the open-loop OCP~\eqref{eq:PCE_SOCP} is in fact a feedback policy with respect to future disturbance realizations. Moreover, the optimal input is not necessarily an affine feedback as long as it lives in the space spanned by the PCE basis $\Psi=\{\psi^j\}_{j=0}^{L-1}$.
\section{The \packname package} \label{sec:package}
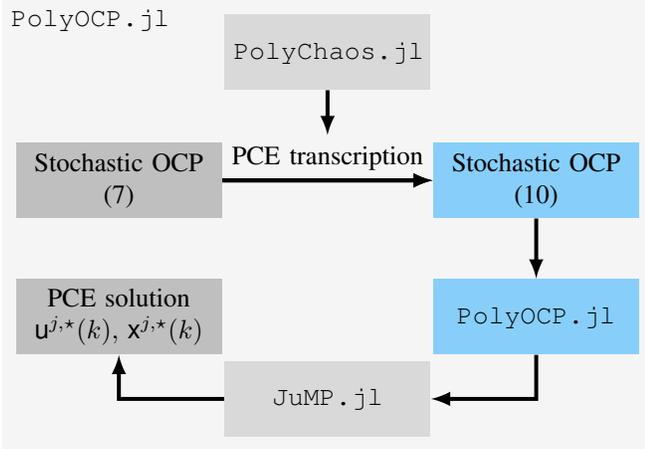
\begin{figure}[t]
    \centering
    \begin{tikzpicture}[auto, >=latex]

      
	\node[fill=lightgray, draw=none, text width = 2.5cm, minimum height=1cm, align=center] (problem1) {Stochastic OCP \\ \eqref{eq:stochasticOCP}};
	\node[fill=lightskyblue, draw=none, text width = 2.5cm, minimum height=1cm, align=center, right = 2.8cm of problem1] (problem2) {Stochastic OCP \\in PCE \eqref{eq:PCE_SOCP}};
	\draw[->, line width=1.5pt] (problem1.east) -- (problem2.west);
	
	\draw[->, line width=1.5pt]  (problem1.east) -- node[above, name=pcetext]{PCE transcription} (problem2.west);
	\node[fill=lightgray!60, draw=none, , text width=2.5cm, minimum height=1cm, align=center,  above=1.2cm of $(problem1)!0.5!(problem2)$]  (polychaos) {\texttt{PolyChaos.jl}};
	\draw[->, line width=1.5pt] (polychaos.south) -- (pcetext.north);
	
	\node[fill=lightskyblue, draw=none, text width = 2.5cm, minimum height=1cm, align=center, below=0.8cm of problem2] (packname) {\packname};
	\draw[->, line width=1.5pt] (problem2.south) -- (packname.north);
	
	\node[fill=lightgray!60, draw=none, , text width=2.5cm, minimum height=1cm, align=center,  below=2.4cm of $(problem1)!0.5!(problem2)$]  (jump) {\texttt{JuMP.jl}};
	\draw[->, line width=1.5pt] (packname.south) |- (jump.east);
	
	\node[fill=lightgray, draw=none, text width = 2.5cm, minimum height=1cm, align=center, below=0.8cm of problem1] (solution) {PCE solution  \\ $\pce{u}^{j,\star}(k),\,\pce{x}^{j,\star}(k)$};
	\draw[->, line width=1.5pt] (jump.west) -|  (solution.south);

    \begin{scope}[on background layer]
		\node[fill=lightgray!15, fit={(polychaos)(problem1)(problem2)(packname)(jump)(solution)}, inner xsep=5pt, inner ysep=5pt, name=box, draw=none] {};
		\node[below right = 0 and 0 of box.north west] {\packname};
	\end{scope}
\end{tikzpicture}
    \caption{Solving stochastic OCPs using \packname}
    \label{fig:flowchart}
\end{figure}

With \packname, we provide a software package written in the Julia programming language that offers a hands-on tool for solving stochastic OCPs within the PCE framework with minimal effort. The workflow of using \packname is illustrated in Fig.~\ref{fig:flowchart}.

\subsection{Workflow in \textnormal{\packname}}
Getting PCE representations of random variables serves as the first step in solving stochastic OCPs within the PCE framework, i.e.~\eqref{eq:PCE_SOCP}. Building upon \code{PolyChaos.jl}, \packname defines the canonical parametric measures and their corresponding orthonormal polynomial functions, as summarized in Table~\ref{tab:PCEBasis}. Note that the use of orthonormal bases in \packname simplifies the formulation of the objective~\eqref{eq:OCP_obj_PCE} and the chance constraint~\eqref{eq:OCP_chance_PCE}, and also improves numerical stability.
Given a random vector $Z\in\lsp({\R^{\dimz})}$, whose components follow the distributions listed in Table~\ref{tab:PCEBasis}, the function \code{genPCE} directly returns the sparse PCE representation of $Z$, including the multivariate orthonormal basis (\code{MultiOrthonoPoly}) and the associated coefficients. Moreover, \packname supports user-defined PCE of $Z$ with multivariate bases up to arbitrary degrees and specified PCE coefficients.

Once the PCEs of $X\ini$ and $W$ are obtained, the function \code{jointPCE} constructs a joint basis for the stochastic LTI system~\eqref{eq:sys} over horizon $N$. In addition, it computes the PCE coefficients of $X\ini$ and $W(k)$, $k\in\I_{[0,N-1]}$, in the joint basis in sparse form.

The data structure for defining \eqref{eq:PCE_SOCP} is a struct called \code{StochOCP} whose fields are described in Table~\ref{tab:StochOCP}. The Boolean flag \code{gauss} determines how $\gamma(\varepsilon)$ is selected in the chance constraint~\eqref{eq:OCP_chance_PCE} as discussed following \eqref{eq:PCE_SOCP}.
Note that chance constraints in \packname are defined on one side as $\mbb P[Z\leq\code{bound}] \geq 1-\code{risk}$, where setting $\code{bound=Inf}$ effectively deactivates the constraint.

Given the required and optional fields, the function \code{buildOCP} automatically generates the optimization problem corresponding to ~\eqref{eq:PCE_SOCP} using a user-specified solver, e.g. \code{Ipopt}, through \code{JuMP.jl}, which is a domain-specific model language for expressing and solving mathematical optimization problems \cite{dunning17jump,waechter06implementation}. Finally, the function \code{solveOCP} solves ~\eqref{eq:PCE_SOCP} and returns the optimal solutions in PCE coefficients $\pce{x}^{j,\star}$ and $\pce{u}^{j,\star}$ for $j\in\I_{[0,L-1]}$.

For using \packname in Model Predictive Control (MPC), the function \code{con\_initial\_param} sets up a constraint for the initial condition with the parametric variable \code{x0param} to avoid rebuilding the entire model repeatedly. The value of \code{x0param} is then updated at each time step with the current measured state using \code{update\_initial\_param} in the MPC loop.
\begin{table}[t]
\caption{Structure \texttt{StochOCP} defining OCP~\eqref{eq:PCE_SOCP}}
\label{tab:StochOCP}
\centering
\renewcommand{\arraystretch}{1.15}
\scalebox{0.92}{
\begin{tabular}{ll}
\hline
\textbf{Field} & \textbf{Description} \\
\hline
\multicolumn{2}{l}{\textit{Required parameters}} \\
\code{N} & Prediction horizon \\
\code{A}, \code{B}, \code{E} & System matrices in~\eqref{eq:OCP_dyn_PCE} \\
\code{x0coeff}, \code{wcoeff} & PCE coefficients of $X_{\mathrm{ini}}$ and $W$ \\[4pt]
\multicolumn{2}{l}{\textit{Optional parameters}} \\
\code{Q}, \code{R}, \code{QN} & Weighting matrices in objective~\eqref{eq:OCP_obj_PCE} \\
\code{lbx}, \code{ubx} & Tuples \code{(bound, risk)} for lower/upper state \\[-1pt]
 & chance constraints; \code{bound} sets $x_{\min}$/$x_{\max}$, \\[-1pt]
 & \code{risk} is possibility of violation allowance \\
\code{lbu}, \code{ubu} & Tuples for input chance constraints (same structure) \\
\code{gauss} & Boolean flag; true only if $X_0$ and $W$ are Gaussians \\
\hline
\end{tabular}
}
\end{table}

\subsection{Additional functionalities of \textnormal{\packname}}
Except for defining~\eqref{eq:PCE_SOCP} through the provided struct \code{StochOCP}, \packname also offers additional functions to enhance modeling flexibility.

When objectives beyond quadratic costs---such as risk-aware objectives---are considered, \packname allows users to specify custom objective functions by leaving the weighting matrix fields in \code{StochOCP} empty. If no objective is defined, \code{JuMP.jl} typically returns a feasible solution.

When only a subset of state or input components is subject to chance constraints, the fields \code{lbx}, \code{ubx}, \code{lbu}, and \code{ubu} may be omitted. Instead, chance constraints can be conveniently imposed using the function \code{con\_chance} that provides an interface for specifying individual constraints. Furthermore, after building the optimization problem via \code{buildOCP}, additional constraints can also be appended following the standard syntax of \code{JuMP.jl}.

To compute the Probability Density Function (PDF) of a random variable from its PCE $Z = \sum_{j=0}^{L-1}\pce{z}^j\psi^j(\xi)\in\lsp(\R)$ numerically, one can use the function \code{pdfPCE} that employs the Fourier transformation $\mcl{F}$ and its inverse $\mcl{F}^{-1}$ as
\begin{equation} \label{eq:PDF}
    f_Z(z) = \mcl{F}^{-1}\left(\textstyle\prod_{j=0}^{L-1}\mcl{F}(\psi^j(\xi))\right).
\end{equation}
The Fourier transforms of the canonical distributions listed in Table~\ref{tab:PCEBasis} are also known as characteristic functions~\cite{oberhettinger73fourier}. 
\section{Illustrative Examples} \label{sec:examples}
We consider two examples to illustrate the features of \packname. The first example demonstrates how to structure a stochastic OCP, whereas the second applies \packname to stochastic MPC. The code for all examples is available in \packname, under the directory~\code{/examples} \cite{polyocp}. All the computations are done in \code{julia} using solver \code{Ipopt} on an AMD Ryzen 9 3900X 12-Core Processor with 3.79 GHz, 64 GB of RAM.

\subsection{Chemical reactor -- Stochastic OCP} \label{ex1:reactor}
We consider the linearized and discretized Van de Vusse reactor model from \cite{heirung18stochastic}. The system matrices are
\[
A = \left[\begin{smallmatrix*}[l] 0.95123 & 0 \\ 0.08833 & 0.81873 \end{smallmatrix*}\right],\,
B = \left[\begin{smallmatrix} -0.0048771 \\-0.0020429 \end{smallmatrix}\right],\,
E = \left[\begin{smallmatrix} 1 \\ 1 \end{smallmatrix}\right].
\]
The initial condition $\Xini$ follows a 2-variate Gaussian distribution as $X\ini\sim\mcl{N}([0.5,\,0.1]^\top,\mathrm{Diag}([0.05^2,\,0.01^2]))$, and disturbances $W$ are uniformly distributed on $[-0.2076,\,0.2076]$. As suggested in Table~\ref{tab:PCEBasis}, a 2-variate Hermite polynomial basis $\{\psi(\xi\ini)\}_{i=0}^2=\{1, \xi_{\tini,1}, \xi_{\tini,2}\}$ is used for $X\ini$, where $\xi_{\tini,1}$ and $\xi_{\tini,2}$ are independent standard Gaussian variables. For $W_k$, $k\in\I_{[0,N-1]}$, Jacobi polynomials
$\{\psi_w^n(\xi_k)\}_{n=0}^1 = \{1,\,2\sqrt{3}(\xi_k-0.5)\}$
are employed, where $\xi_k$ are i.i.d. variables following $\mcl{U}(0,1)$. Their PCEs and the joint basis from~\eqref{eq:jointbasis} are then given by
\begin{align*}
	 X\ini &= \begin{bmatrix} 0.5 \\ 0.1 \end{bmatrix} \cdot 1 + 
		 \begin{bmatrix*}[l] 0.05 \\ 0 \end{bmatrix*} \cdot \psi^1(\xi\ini) + 
		 \begin{bmatrix*}[l] 0 \\ 0.01 \end{bmatrix*} \cdot \psi^2(\xi\ini),\\
    W_k &= 0\cdot 1 + \frac{2\cdot 0.2076}{2\sqrt{3}}\psi^1(\xi_k),\quad k\in\I_{[0,N-1]},\\
    \Psi &= \left\{1, \psi^1(\xi\ini), \psi^2(\xi\ini), \psi^1(\xi_0),\cdots,\psi^1(\xi_{N-1})\right\},
\end{align*}
where $\Psi$ consists of $N+3$ terms. The prediction horizon is set to $N=50$ and weighting matrices are chosen as identical matrices, i.e. $Q=Q_N=I_2$ and $R=1$.

\begin{figure}[t]
	\begin{center}
		\includegraphics[width=0.9\linewidth,trim={68mm 13mm 50mm 25mm},clip]{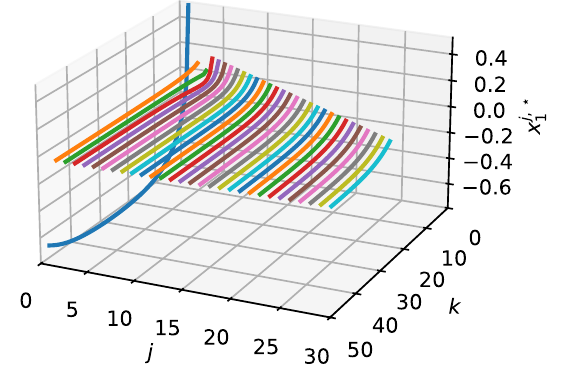}
		\caption{Trajectories of the first 30 PCE coefficients of $X_1$ for the chemical reactor} \label{fig:EX1_PCEX1}	
	\end{center}
\end{figure}

\begin{figure}[t]
	\begin{center}
        \includegraphics[width=0.9\linewidth,trim={55mm 10mm 60mm 26mm},clip]{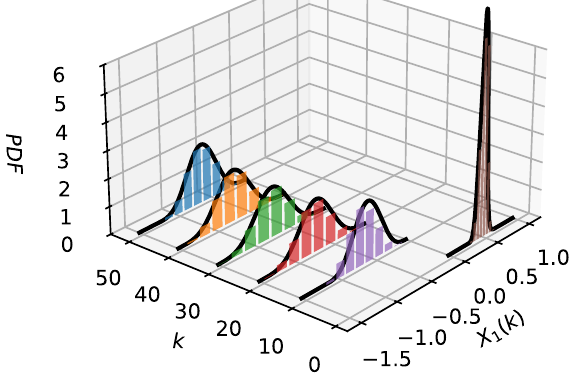}
		\caption{Comparison of PDFs and histograms of $10^4$ samples for the chemical reactor} \label{fig:EX1_DistributionX1}	
	\end{center}
\end{figure}

To solve the stochastic OCP, a struct~\code{StochOCP} is defined with the given parameters. In addition, a state chance constraint $\mbb P[X_2(k)\leq0.24] \geq 0.9$, $k\in\I_{[0,N]}$ is imposed by letting \code{StochOCP.ubx = ([Inf;0.24], [1;0.1])}, where the upper bound value \texttt{Inf} indicates that the constraint on $X_1$ is deactivated. After building the PCE OCP with 8056 decision variables via \code{buildOCP}, \code{solveOCP} directly returns the optimal solution in terms of PCE coefficients.
The average computation time for solving the OCP over 1000 runs is 159.68~\unit{\milli\second}.
The trajectories of the first 30 PCE coefficients of the state component $X_1$, i.e. $\pce{x}_1^j$ for $j=\I_{[0,29]}$, are depicted in Fig.~\ref{fig:EX1_PCEX1}. We observe that the PCE coefficients exhibit a triangular structure due to the causality constraint~\eqref{eq:OCP_chance_PCE}, and we refer  to \cite{ou25polynomial} for an in-depth analysis.
By drawing $10^4$ samples of $\xi\ini$ and $\xi_k$, $k\in\I_{[0,N-1]}$, for PCE basis, we obtain the histograms of $X_1(k)$ at time steps $k=0,10,20,30,40,50$. These histograms perfectly match the corresponding PDFs calculated from~\eqref{eq:PDF} as shown in Fig. \ref{fig:EX1_DistributionX1}. The code for this example is provided in \code{examples/ChemicalReactor.jl}.

\subsection{Four tank system -- Stochastic MPC}
We consider a linearized version of a four tank system taken from \cite{berberich20data} with system matrices
\[
A =
\left[\begin{smallmatrix*}[l]
0.921 & 0     & 0.041 & 0 \\
0     & 0.918 & 0     & 0.033 \\
0     & 0     & 0.924 & 0 \\
0     & 0     & 0     & 0.937
\end{smallmatrix*}\right], \,
B =
\left[\begin{smallmatrix*}[l]
0.017 & 0.001 \\
0.001 & 0.023 \\
0     & 0.061 \\
0.072 & 0
\end{smallmatrix*}\right],
\]
and $E=I_4$. Each component of $W$ is assumed to be independent with distribution $W_i\sim 0.05\cdot(Z^2+Z)$, where $Z\sim\mcl{N}(0,1)$ is a standard Gaussian variable. Hence, Hermite polynomials up to degree two are chosen for each component $W_i$, whose PCE reads
\[
    W_i = 0.05\cdot\Big(1\cdot 1 + 1\cdot \psi^1(\xi) + \sqrt{2}\cdot\psi^2(\xi)\Big),\quad \xi\sim\mcl{N}(0,1),
\]
where $\psi^1(\xi)=\xi$ and $\psi^2=(\xi^2-1)/\sqrt{2}$ are listed in Table~\ref{tab:PCEBasis}. Each component of the initial state $X(0)$ is sampled uniformly from $[-1,1]$ to generate different initial conditions. Moreover, the weighting matrices are chosen to be $Q=Q_N=3\cdot I_4$ and $R=10^4\cdot I_2$. The chance constraints are imposed on the first two components of the state as $\mbb P[-2\leq X_i \leq 2] \geq 0.8$, $i=1,2$. Similar to Example~\ref{ex1:reactor}, the corresponding \code{StochOCP} with prediction horizon $N=10$ is constructed and the model is built. Note that when running \packname in MPC, the parametric initial condition \code{x0param} is set to the current measured state. Further details about this example are provided in~\code{examples/Tank.jl}.

\begin{figure}[t]
	\begin{center}
        \includegraphics[width=1\linewidth]{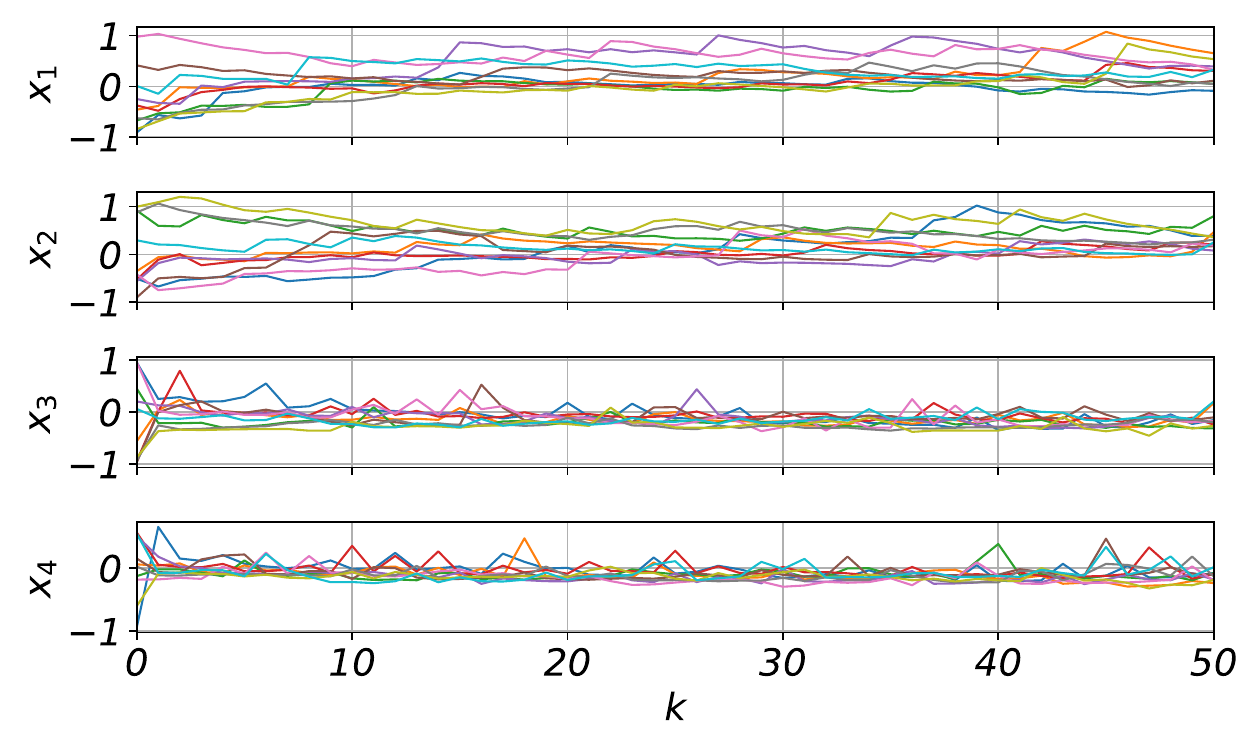}
		\caption{10 different closed-loop realizations of state trajectories for the four tank system}
        \label{fig:EX2_TankSampleTraj}	
	\end{center}
\end{figure}

\begin{figure}
	\begin{subfigure}{\linewidth}
		\centering
		\includegraphics[width=0.9\linewidth,trim={55mm 10mm 60mm 28mm},clip]{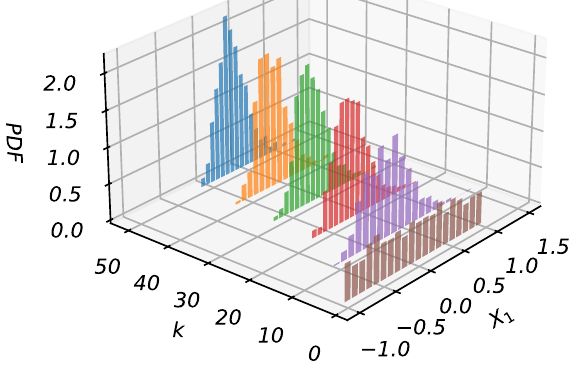}
	\end{subfigure}
	\begin{subfigure}{\linewidth}
		\centering
		\includegraphics[width=0.9\linewidth,trim={55mm 10mm 60mm 28mm},clip]{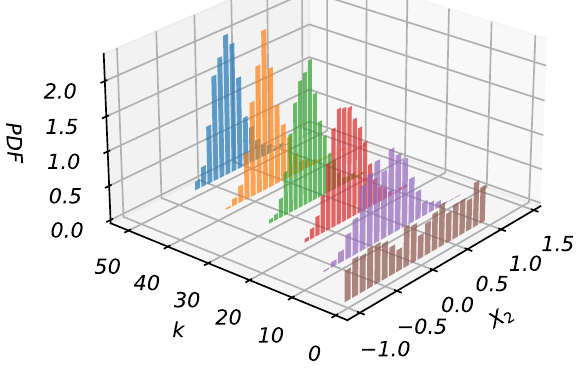}
	\end{subfigure}\\[1ex]
	\caption{Time evolution of empirical distributions of  $X_i$, $i=1,2$ of the closed-loop four tank system}
	\label{fig:EX2_TankDistribution}
\end{figure}

We sample 1000 initial states of the four tank system and conduct a closed-loop simulation over 50 steps for each, resulting in a total of 50.000 stochastic OCPs. At each time step, a stochastic OCP in PCE~\eqref{eq:PCE_SOCP} is solved with the measured state as deterministic initial condition, while the disturbances are modeled as random variables. The state evolution is obtained using independently sampled disturbance paths for each initial state.
10 sampled closed-loop realization trajectories are shown in Fig~\ref{fig:EX2_TankSampleTraj}. It can be seen that the state realizations stay close to the origin over time. In addition, the empirical distributions of the closed-loop state trajectories $X_i$, $i=1,2$, are depicted in Fig.~\ref{fig:EX2_TankDistribution}. Observe that all the realizations of $X_i$, $i=1,2$ lie in the interval~$[-2,2]$, since the chance constraint reformulation~\eqref{eq:OCP_chance_PCE} is conservative. Moreover, the realization paths are sampled using both parallel and serial methods in \code{julia}, and the computation times are reported in Table~\ref{tab:EX2_times}. The column “Per sample” denotes the average time for one closed-loop trajectory of 50 steps, and “Per OCP” the average solution time of a single stochastic OCP. In the parallel method, the samples are computed simultaneously on 10~cores, resulting in an 82.01\% reduction in total computation time. On average, solving a single stochastic OCP using PCE with 5184 decision variables and linear and second-order cone constraints, takes 10.50~\unit{\milli\second} with parallelized sampling.

Further examples, e.g., stochastic LTI systems with non-i.i.d. additive disturbances, are available under \code{/examples}.

\begin{table}[!t]
\caption{Comparison of computation times for parallel and serial sampling methods for the four tank system.}
\label{tab:EX2_times}
\centering
\renewcommand{\arraystretch}{1.2}
\begin{tabular}{cccc}
\hline
Method & Overall time (\unit{\second}) & Per sample (\unit{\milli\second}) & Per OCP (\unit{\milli\second}) \\ 
\hline
Parallel  & \phantom{2}524.90 & \phantom{2}524.91 & 10.50 \\ 
Serial   & 2919.31 & 2919.31 & 58.39 \\ 
\hline
\end{tabular}
\end{table}
\section{Conclusion} \label{sec:conclusions}
This paper has introduced \packname, a \code{julia} toolbox, released under the MIT license, for solving stochastic optimal control problems in the PCE framework. Existing PCE tools are not tailored to handle problems with additive stochastic disturbances. \packname presents a first step towards closing this gap.  The toolbox is intended to facilitate rapid prototyping of stochastic OCPs and corresponding stochastic MPC schemes.

Future work will consider the inclusion of data-driven OCP formulations~\cite{pan23stochastic} and the extension towards distributionally robust formulations~\cite{pan23distributionally}. Also the consideration of risk-sensitive objective and constraint formulations (beyond chance constraints) will be of interest. 

\addtolength{\textheight}{0cm}   



%

\section*{ACKNOWLEDGMENT}
During the preparation of this work the authors used ChatGPT in order to  check grammar. After using this tool, the authors reviewed and edited the content as needed and take full responsibility for the content of the published article.

\bibliographystyle{IEEEtran}
\bibliography{arXivRef}

\begin{thebibliography}{10}
\providecommand{\url}[1]{#1}
\csname url@samestyle\endcsname
\providecommand{\newblock}{\relax}
\providecommand{\bibinfo}[2]{#2}
\providecommand{\BIBentrySTDinterwordspacing}{\spaceskip=0pt\relax}
\providecommand{\BIBentryALTinterwordstretchfactor}{4}
\providecommand{\BIBentryALTinterwordspacing}{\spaceskip=\fontdimen2\font plus
\BIBentryALTinterwordstretchfactor\fontdimen3\font minus
  \fontdimen4\font\relax}
\providecommand{\BIBforeignlanguage}[2]{{%
\expandafter\ifx\csname l@#1\endcsname\relax
\typeout{** WARNING: IEEEtran.bst: No hyphenation pattern has been}%
\typeout{** loaded for the language `#1'. Using the pattern for}%
\typeout{** the default language instead.}%
\else
\language=\csname l@#1\endcsname
\fi
#2}}
\providecommand{\BIBdecl}{\relax}
\BIBdecl

\bibitem{Witsenhausen71}
H.~Witsenhausen, ``Separation of estimation and control for discrete time
  systems,'' \emph{Proceedings of the IEEE}, vol.~59, no.~11, pp. 1557--1566,
  1971.

\bibitem{Florentin61a}
J.~J. Florentin, ``Optimal control of continuous time {Markov} stochastic
  systems,'' \emph{International Journal of Electronics}, vol.~10, no.~6, pp.
  473--488, 1961.

\bibitem{kushner1964maximum}
H.~J. Kushner and F.~C. Schweppe, ``A maximum principle for stochastic control
  systems,'' \emph{Journal of Mathematical Analysis and Applications}, vol.~8,
  no.~2, pp. 287--302, 1964.

\bibitem{farina13probabilistic}
M.~Farina, L.~Giulioni, L.~Magni, and R.~Scattolini, ``A probabilistic approach
  to model predictive control,'' in \emph{52nd IEEE Conference on Decision and
  Control (CDC)}.\hskip 1em plus 0.5em minus 0.4em\relax IEEE, 2013, pp.
  7734--7739.

\bibitem{fagiano12nonlinear}
L.~Fagiano and M.~Khammash, ``Nonlinear stochastic model predictive control via
  regularized polynomial chaos expansions,'' in \emph{51st IEEE Conference on
  Decision and Control (CDC)}.\hskip 1em plus 0.5em minus 0.4em\relax IEEE,
  2012, pp. 142--147.

\bibitem{hewing18a}
L.~Hewing and M.~N. Zeilinger, ``Stochastic model predictive control for linear
  systems using probabilistic reachable sets,'' in \emph{2018 IEEE Conference
  on Decision and Control (CDC)}.\hskip 1em plus 0.5em minus 0.4em\relax IEEE,
  2018, pp. 5182--5188.

\bibitem{mcallister22a}
R.~D. McAllister and J.~B. Rawlings, ``Nonlinear stochastic model predictive
  control: Existence, measurability, and stochastic asymptotic stability,''
  \emph{IEEE Transactions on Automatic Control}, vol.~68, no.~3, pp.
  1524--1536, 2022.

\bibitem{schiessl25stability}
J.~Schie{\ss}l, , H.~Selder, R.~Ou, T.~Faulwasser, M.~H. Baumann, and
  L.~Gr{\"u}ne, ``Stability and performance of stochastic economic {MPC} -
  stochastic characterization of the closed-loop asymptotics,''
  \emph{arXiv:2510.19416}, 2025.

\bibitem{calafiore06a}
G.~C. Calafiore and M.~C. Campi, ``The scenario approach to robust control
  design,'' \emph{IEEE Transactions on Automatic Control}, vol.~51, no.~5, pp.
  742--753, 2006.

\bibitem{athans71a}
M.~Athans, ``The role and use of the stochastic linear-quadratic-gaussian
  problem in control system design,'' \emph{IEEE Transactions on Automatic
  Control}, vol.~16, no.~6, pp. 529--552, 1971.

\bibitem{landgraf23a}
D.~Landgraf, A.~V{\"o}lz, F.~Berkel, K.~Schmidt, T.~Specker, and K.~Graichen,
  ``Probabilistic prediction methods for nonlinear systems with application to
  stochastic model predictive control,'' \emph{Annual Reviews in Control},
  vol.~56, p. 100905, 2023.

\bibitem{schiessl25towards}
J.~Schie{\ss}l, R.~Ou, M.~H. Baumann, T.~Faulwasser, and L.~Gr{\"u}ne,
  ``Towards turnpike-based performance analysis of risk-averse stochastic
  predictive control,'' \emph{ArXiv:2504.00701}, 2025.

\bibitem{pan25dissertation}
G.~Pan, ``Data-driven control of stochastic systems: Representation,
  prediction, and optimal control,'' Ph.D. dissertation, Institute of Control
  Systems, Hamburg University of Technology (TUHH), 2025.

\bibitem{sullivan15introduction}
T.~Sullivan, \emph{{Introduction to Uncertainty Quantification}}.\hskip 1em
  plus 0.5em minus 0.4em\relax Springer International, 2015, vol.~63.

\bibitem{wiener38homogeneous}
N.~Wiener, ``The homogeneous chaos,'' \emph{American Journal of Mathematics},
  pp. 897--936, 1938.

\bibitem{kim12probabilistic}
K.-K.~K. Kim and R.~D. Braatz, ``Probabilistic analysis and control of
  uncertain dynamic systems: Generalized polynomial chaos expansion
  approaches,'' in \emph{2012 American Control Conference}, 2012, pp. 44--49.

\bibitem{kim13wiener}
K.-K.~K. Kim, D.~E. Shen, Z.~K. Nagy, and R.~D. Braatz, ``Wiener's polynomial
  chaos for the analysis and control of nonlinear dynamical systems with
  probabilistic uncertainties [{H}istorical {P}erspectives],'' \emph{IEEE
  Control Systems Magazine}, vol.~33, no.~5, pp. 58--67, 2013.

\bibitem{kim13generalised}
K.-K.~K. Kim and R.~D. Braatz, ``Generalised polynomial chaos expansion
  approaches to approximate stochastic model predictive control,''
  \emph{International journal of control}, vol.~86, no.~8, pp. 1324--1337,
  2013.

\bibitem{mishra24a}
P.~K. Mishra, J.~A. Paulson, and R.~D. Braatz, ``Polynomial chaos-based
  stochastic model predictive control: An overview and future research
  directions,'' \emph{arXiv:2406.10734}, 2024.

\bibitem{petzke20pocet}
F.~Petzke, A.~Mesbah, and S.~Streif, ``Pocet: a polynomial chaos expansion
  toolbox for matlab,'' \emph{IFAC-PapersOnLine}, vol.~53, no.~2, pp.
  7256--7261, 2020, 21st IFAC World Congress.

\bibitem{muehlpfordt20polychaos}
T.~M{\"u}hlpfordt, F.~Zahn, V.~Hagenmeyer, and T.~Faulwasser,
  ``{P}oly{C}haos.jl—{A} {J}ulia package for polynomial chaos in systems and
  control,'' \emph{IFAC-PapersOnLine}, vol.~53, no.~2, pp. 7210--7216, 2020,
  21th IFAC World Congress.

\bibitem{ou25polynomial}
R.~Ou, J.~Schie{\ss}l, M.~H. Baumann, L.~Gr{\"u}ne, and T.~Faulwasser, ``A
  polynomial chaos approach to stochastic {LQ} optimal control: Error bounds
  and infinite-horizon results,'' \emph{Automatica}, vol. 174, p. 112117, 2025.

\bibitem{polyocp}
``\texttt{PolyOCP.jl},'' \url{https://github.com/OptCon/PolyOCP.jl}, 2026,
  version 0.1.2.

\bibitem{dunning17jump}
I.~Dunning, J.~Huchette, and M.~Lubin, ``{JuMP}: A modeling language for
  mathematical optimization,'' \emph{SIAM Review}, vol.~59, no.~2, pp.
  295--320, 2017.

\bibitem{Kapernick14a}
B.~K{\"a}pernick and K.~Graichen, ``The gradient based nonlinear model
  predictive control software {GRAMPC},'' in \emph{2014 European Control
  Conference (ECC)}.\hskip 1em plus 0.5em minus 0.4em\relax IEEE, 2014, pp.
  1170--1175.

\bibitem{Andersson19a}
J.~Andersson, J.~Gillis, G.~Horn, J.~Rawlings, and M.~Diehl, ``Casadi: {A}
  software framework for nonlinear optimization and optimal control,''
  \emph{Mathematical Programming Computation}, vol.~11, no.~1, pp. 1--36, 2019.

\bibitem{pan23distributionally}
G.~Pan and T.~Faulwasser, ``Distributionally robust uncertainty quantification
  via data-driven stochastic optimal control,'' \emph{IEEE Control Systems
  Letters}, vol.~7, pp. 3036--3041, 2023.

\bibitem{feinberg15chaospy}
J.~Feinberg and H.~P. Langtangen, ``Chaospy: An open source tool for designing
  methods of uncertainty quantification,'' \emph{Journal of Computational
  Science}, vol.~11, pp. 46--57, 2015.

\bibitem{marelli14uqlab}
S.~Marelli and B.~Sudret, ``{UQLab}: {A} framework for uncertainty
  quantification in {MATLAB},'' in \emph{The 2nd International Conference on
  Vulnerability and Risk Analysis and Management (ICVRAM 2014)}, 2014, pp.
  2554--2563.

\bibitem{baudin17openturns}
M.~Baudin, A.~Dutfoy, B.~Iooss, and A.-L. Popelin, \emph{OpenTURNS: An
  Industrial Software for Uncertainty Quantification in Simulation}.\hskip 1em
  plus 0.5em minus 0.4em\relax Cham: Springer International Publishing, 2017,
  pp. 2001--2038.

\bibitem{adams23dakota}
B.~M. Adams, W.~J. Bohnhoff, K.~R. Dalbey \emph{et~al.}, ``Dakota 6.19.0
  documentation,'' Sandia National Laboratories, Albuquerque, NM, Tech. Rep.
  SAND2023-13392O, 2023.

\bibitem{cameron47orthogonal}
R.~Cameron and W.~Martin, ``The orthogonal development of non-linear
  functionals in series of {F}ourier-{H}ermite functionals,'' \emph{Annals of
  Mathematics}, pp. 385--392, 1947.

\bibitem{ernst12convergence}
O.~Ernst, A.~Mugler, H.-J. Starkloff, and E.~Ullmann, ``On the convergence of
  generalized polynomial chaos expansions,'' \emph{ESAIM: Mathematical
  Modelling and Numerical Analysis}, vol.~46, no.~2, pp. 317--339, 2012.

\bibitem{xiu02wiener}
D.~Xiu and G.~Karniadakis, ``The {W}iener--{A}skey polynomial chaos for
  stochastic differential equations,'' \emph{SIAM Journal on Scientific
  Computing}, vol.~24, no.~2, pp. 619--644, 2002.

\bibitem{fristedt13modern}
B.~Fristedt and L.~Gray, \emph{A Modern Approach to Probability Theory}.\hskip
  1em plus 0.5em minus 0.4em\relax Birkh{\"a}user Boston, 1997.

\bibitem{pan23stochastic}
G.~Pan, R.~Ou, and T.~Faulwasser, ``On a stochastic fundamental lemma and its
  use for data-driven optimal control,'' \emph{IEEE Transactions on Automatic
  Control}, vol.~68, no.~10, pp. 5922--5937, 2023.

\bibitem{waechter06implementation}
A.~W{\"a}chter and L.~T. Biegler, ``On the implementation of an interior-point
  filter line-search algorithm for large-scale nonlinear programming,''
  \emph{Mathematical Programming}, vol. 106, no.~1, pp. 25--57, 2006.

\bibitem{oberhettinger73fourier}
F.~Oberhettinger, \emph{Fourier transforms of distributions and their inverses:
  a collection of tables}.\hskip 1em plus 0.5em minus 0.4em\relax Academic
  Press, 1973.

\bibitem{heirung18stochastic}
T.~A.~N. Heirung, J.~A. Paulson, J.~O’Leary, and A.~Mesbah, ``Stochastic
  model predictive control—how does it work?'' \emph{Computers \& Chemical
  Engineering}, vol. 114, pp. 158--170, 2018.

\bibitem{berberich20data}
J.~Berberich, J.~K{\"o}hler, M.~A. M{\"u}ller, and F.~Allg{\"o}wer,
  ``Data-driven model predictive control with stability and robustness
  guarantees,'' \emph{IEEE Transactions on Automatic Control}, vol.~66, no.~4,
  pp. 1702--1717, 2020.

\end{thebibliography}

\end{document}